# Interactive Treatment Planning in High Dose-Rate Brachytherapy for Gynecological Cancer

Huan Liu[1], Chang M Ma[1†], Xun Jia[2†], Chenyang Shen[2], Peter Klages[2], Kevin Albuquerque[2]

[1]Department of Radiation Oncology, Fox Chase Cancer Center, Philadelphia, PA 19111, USA

[2]Department of Radiation Oncology, University of Texas Southwestern Medical Center, Dallas, TX 75287, USA

†Corresponding Author: Charlie.Ma@fccc.edu and xun.jia@utsouthwestern.edu

**Abstract**

Purpose: High dose-rate brachytherapy (HDRBT) is widely used for gynecological cancer treatment. Although commercial treatment planning systems (TPSs) have inverse optimization modules, it takes several iterations to adjust planning objectives to achieve a satisfactory plan. Interactive plan-modification modules enable modifying the plan and visualizing results in real time, but they update plans based on simple geometrical or heuristic algorithms, which cannot ensure resulting plan optimality. This project develops an interactive plan optimization module for HDRBT of gynecological cancer. By efficiently solving an optimization problem in real time, it allows a user to visualize a plan and interactively modify it to improve quality.

Methods: We formulated an optimization problem with an objective function containing a weighted sum of doses to normal organs subject to user-specified target coverage. A user interface was developed that allows a user to adjust organ weights using scroll bars. With a simple mouse click, the optimization problem is solved in seconds with a highly efficient alternating-direction method of multipliers and a warm start optimization strategy. Resulting clinically relevant $D_{2cc}$ of organs are displayed immediately. This allows a user to intuitively adjust plans with satisfactory quality. We tested the effectiveness of our development in cervix cancer cases treated with a tandem-and-ovoid applicator.

Results: It took a maximum of 3 seconds to solve the optimization problem in each instance. With interactive optimization capability, a satisfactory plan can be obtained in <1 min. In our clinic, although the time for plan adjustment was typically <10min with simple interactive plan modification tools in TPS, the resulting plans do not ensure optimality. Our plans achieved on average 5% lower $D_{2cc}$ than clinical plans, while maintaining target coverage.

Conclusion: The interactive optimization tool is effective in terms of helping a planner to generate a plan with improved efficiency and plan quality.

## 1 Introduction

High dose-rate brachytherapy (HDRBT) is an important radiotherapy modality for the management of gynecological cancer. It delivers radiation by placing an applicator in the target area and using a motor-controlled radioactive source to dwell in a series of planned positions with preset times. The highly conformal dose distribution has clearly demonstrated its clinical advantages, such as better local control[1] and improved cancer cure rates[2].

Treatment planning is a crucial step for the success of HDRBT, which critically affects the plan quality. In a typical clinical setting, the planner is required to complete the treatment planning process in a short time frame[3] in a high-stress environment. Hence, it is of central importance to have an effective treatment-planning tool for the planner to quickly navigate the solution space and obtain the appropriate solution meeting the physician's requirements.

There are currently two main approaches in existing commercial treatment planning systems (TPSs). The first is inverse treatment plan optimization. The treatment-planning problem is formulated as an optimization problem. The planner can specify planning objectives through an interface and launch the optimization process. After assessing the resulting solution quality with the physician, the planner makes decisions about modifying the planning objectives and re-launches the optimization process. This process continues until a satisfactory plan is obtained. This approach ensures Pareto optimality of the resulting plan [4]. Nonetheless, due to the iterative nature and the relatively lengthy period of time to solve the optimization problem each time, the overall process is time-consuming. This is further exacerbated by the fact that the physician typically cannot wait for the planner to adjust planning objectives due to his busy schedule, and hence, leaves the planning room frequently. As a consequence, the additional time of locating the physician to consult about the plan quality further slows the process. The second typical approach in current TPS is to interactively modify the plan and observe resulting plans immediately. Due to computational challenges, the plan modification is implemented by simple and non-optimization based approaches. For instance, the dose shaper tool in the BrachyVision system (Varian Medical Systems, Palo Alto, CA) allows the planner to drag isodose lines. The dwell time is updated according to geometrical distances between mouse positions and the dwell positions. While this is simple and allows the user to see changes in real time, Pareto optimality of the resulting plan cannot be guaranteed. We recently conducted a retrospective analysis of 96 plans for cervical cancer patients treated with HDRBT using a tandem-and-ovoid applicator. It was found that the clinically delivered plans, which were obtained using the dose shaper tool, can be further improved. On average, if inverse optimization were carefully performed, $D_{2cc}$ can be further reduced for the bladder, rectum, and sigmoid by 5.41%, 14.7%, and 5.21% and the maximum reductions were 47.1%, 59.9%, and 45.9%, respectively[5], without compromising target coverage. As the $D_{2cc}$ values are indicators of treatment toxicity[6], the reductions translate into clinical benefits in terms of average reduction of toxicity rates of bladder, rectum, and sigmoid by 3.81%, 6.15%, and 5.27%, and the maximum reduction 24.53%, 58.22%, and 69.54%, respectively.

In light of the need for an effective treatment-planning tool to quickly generate high-quality treatment plans, a pioneer study was conducted to develop an interactive multi-objective optimization approach[7]. It employed an existing optimization tool, NIMBUS[8], to solve a multi-objective optimization problem in an interactive manner. During the planning process, the planner classifies objective functions of the observed plan and gives preference information about how the current solution should be improved. Based on this preference, a sub-problem is formulated, which is solved with an appropriate optimizer to generate the next solution. While the effectiveness of this method has been demonstrated in two example test cases, the use of NIMBUS restricted the decision making to be based on objective function values. In their study, the objective function value does not clearly link to clinical objectives, such as $D_{2cc}$[9,10]. In addition, it takes several minutes each time to solve the non-convex sub-problem, which deteriorates the interactivity between the planner and the planning system and impedes the overall workflow.

In this paper, we report our development on a new interactive optimization-based planning tool, as part of the AutoBrachy system developed at our institution[11-13]. In this tool, we formulated an optimization problem with organ weights adjustable by the planner via an interface. We further developed an efficient numerical algorithm to solve the optimization problem, such that the results can be obtained in a few seconds and the clinically relevant $D_{2cc}$ values are displayed to facilitate the planner to adjust the parameters. This tool is expected to help the planner to quickly navigate the solution space and reach the targeted high-quality plan efficiently.

## 2 Methods and Materials

### 2.1 Optimization problem

The aim of HDRT treatment planning for gynecological cancer is to obtain a plan with sufficient tumor coverage while maintaining dose to organs at risk (OARs) to an acceptable level. In our clinic, the tumor coverage is quantified by $D_{90}$ (the minimum dose that covers at least 90% volume of the clinical target volume), and the OARs dose is by $D_{2cc}$ (the minimum dose that irradiates a volume of 2 cm$^3$ of the organ of interest). Typical organs considered include bladder, rectum, sigmoid colon, and small bowel.

With this in mind, we consider an optimization problem as

$$t = \underset{t}{\operatorname{argmin}} \sum_i \frac{\lambda_i}{2} \left\| M_{OAR}^i t \right\|_2^2 + \frac{1}{2} \|t\|_2^2, \tag{1}$$

$$s.t. \; D^{CTV} = M_{CTV} t, D^{CST} = M_{CST} t$$

$$^1 \; D_{90}^{CTV} = \alpha p,$$

$$^2 \; D^{CST} \in [0.8p, \; 1.4p],$$

$$t \in [0, \; t_{max}],$$

where $t$ is the dwell time to be determined. $M_{OAR}^i$, $M_{CTV}$, and $M_{CST}$ denote the dose deposition matrices of the ith OARs, to CTV, and to voxels at certain locations that are defined for clinical consideration, respectively. These matrices are calculated based on the Task group report number 43 of AAPM. For those voxels defined for clinical consideration corresponding with $M_{CST}$, two sets of points are labeled. One set of points is in a parallel to the ovoid applicator, and the dose they receive is constrained by physicians to prevent metastasis of tumor to those regions. The other set of points is defined as a surface near the tandem tip to regulate the dose pattern to a pear shape. $\lambda_i$ is the weighting factor of the ith OAR and $\alpha$ is a parameter close to unity, specifying the desired tumor coverage. $p$ is the prescription dose. These parameters are adjusted by the planner to yield a satisfactory plan. This optimization problem contains a box constraint for the dwell time. In addition, the range of dose matrix of empirical constraints is $[0.8\,p, 1.4\,p]$ during the optimization.

To solve this objective function, we employed the Alternating Direction Method of Multipliers. Let $\widehat{M} = \begin{pmatrix} M_{PTV} \\ M_{CST} \end{pmatrix}, x = \begin{pmatrix} D^{PTV} \\ D^{CST} \end{pmatrix}$, the augmented Lagrangian function of the optimization problem is as follows:

$$L(t, x, \Gamma) = \sum_i \frac{\lambda_i}{2} \left\| M_{OAR}^i t \right\|_2^2 + \frac{1}{2} \|t\|_2^2 + \frac{\beta}{2} \left\| \widehat{M} t - x \right\|_2^2 + \langle \Gamma, \widehat{M} t - x \rangle + \delta_1(x) + \delta_{box}(t) \tag{2}$$

For iteration number $k = 0, 1, \ldots, N$, $t$ sub-problem (formula 3, 4) and $x$ sub-problem (formula 5, 6) need to be solved, then update $\Gamma$ using formula 7. A linear solver called the conjugate gradient method is used to solve $t$ sub-problem. When solving the $x$ sub-problem, constraints 1 and 2 need to be ensured. The solving process will be terminated when the number of iterations is greater than $10^3$ or the difference of norm of time vector between two iterations is less than $10^{-3}$.

$$(\sum_i \lambda_i {M_{OAR}^i}^T M_{OAR}^i + I + \beta \widehat{M}^T \widehat{M}) t^{\left(k+\frac{1}{2}\right)} = \beta \widehat{M}^T x^{(k)} - \widehat{M}^T \Gamma^{(k)} \tag{3}$$

$$t^{(k+1)} = \text{proj}_{\text{box}}\left\{t^{\left(k+\frac{1}{2}\right)}\right\}, i.e. \left[t^{(k+1)}\right]_i = \begin{cases} 0, & \text{if } \left[t^{\left(k+\frac{1}{2}\right)}\right]_i < 0 \\ t_{max}, & \text{if } \left[t^{\left(k+\frac{1}{2}\right)}\right]_i > t_{max} \\ \left[t^{\left(k+\frac{1}{2}\right)}\right]_i, & \text{otherwise} \end{cases} \quad (4)$$

$$x^{\left(k+\frac{1}{2}\right)} = \widehat{M}t^{(k+1)} + \frac{\Gamma^{(k)}}{\beta} \quad (5)$$

$$x^{(k+1)} = \text{proj}_{1,2}\left\{x^{\left(k+\frac{1}{2}\right)}\right\} \quad (6)$$

$$\Gamma^{(k+1)} = \Gamma^{(k)} + \beta(\widehat{M}t^{(k+1)} - x^{(k+1)}) \quad (7)$$

*2.2 Interface design and workflow*

We developed the following workflow and interface to integrate the optimization problem into clinical practice. Before the optimization, patient geometries, including structures of target and OARs and dwell positions within the applicator, were generated with BrachyVision. Subsequently, the data is exported to our system via DICOM-RT. To facilitate treatment planning, we developed an interactive optimization interface using C# to interactively optimize the treatment plan.

Before optimization, the dose rate matrix of all selected points should be entered into the program to do the initialization. When the optimization program is started, an interactive optimization interface (IOI), as shown in figure 1, pops up for treatment planner. The left side of the interface is a region for the definitions of organ weighting factors and CTV coverage. The five axes of a regular pentagon represent CTV and four OARs respectively. There are five-line segments connecting the center point of the pentagon with the corresponding vertices. Five sliders are put on the line segments respectively. When the treatment planner defines the weighting factors for CTV and OARs, he simply needs to drag the corresponding sliders. The weighting factor will vary from minimum to maximum when a slider is dragged from center to vertices. In the optimization model, the range of the weighting factors of target and OARs is [0.8,1.2] and [0,0.3] respectively. The slider bar for OARs is in log scale. Moreover, a treatment planner can also directly input the value into the corresponding textbox, making the adjustment more efficient in the situation where a slight variation can cause a big change for optimization results.

Once weighting factors of CTV and OARs are determined, the optimization bar is clicked, and subsequently the right side of the interface presents the current plan with the value of $D_{90}$ for CTV and $D_{2cc}$ for OARs in 2 seconds. The treatment planner checks whether the current optimization plan is satisfactory according to the $D_{2cc}$ value for specified OARs. If the current plan fulfills clinical requirements, the optimization is terminated. Otherwise, the planner adjusts the weighting factors iteratively, until obtaining an optimal plan. Upon completion, the resulting plan is transferred to the clinical TPS BrachyVision via DICOM-RT data.

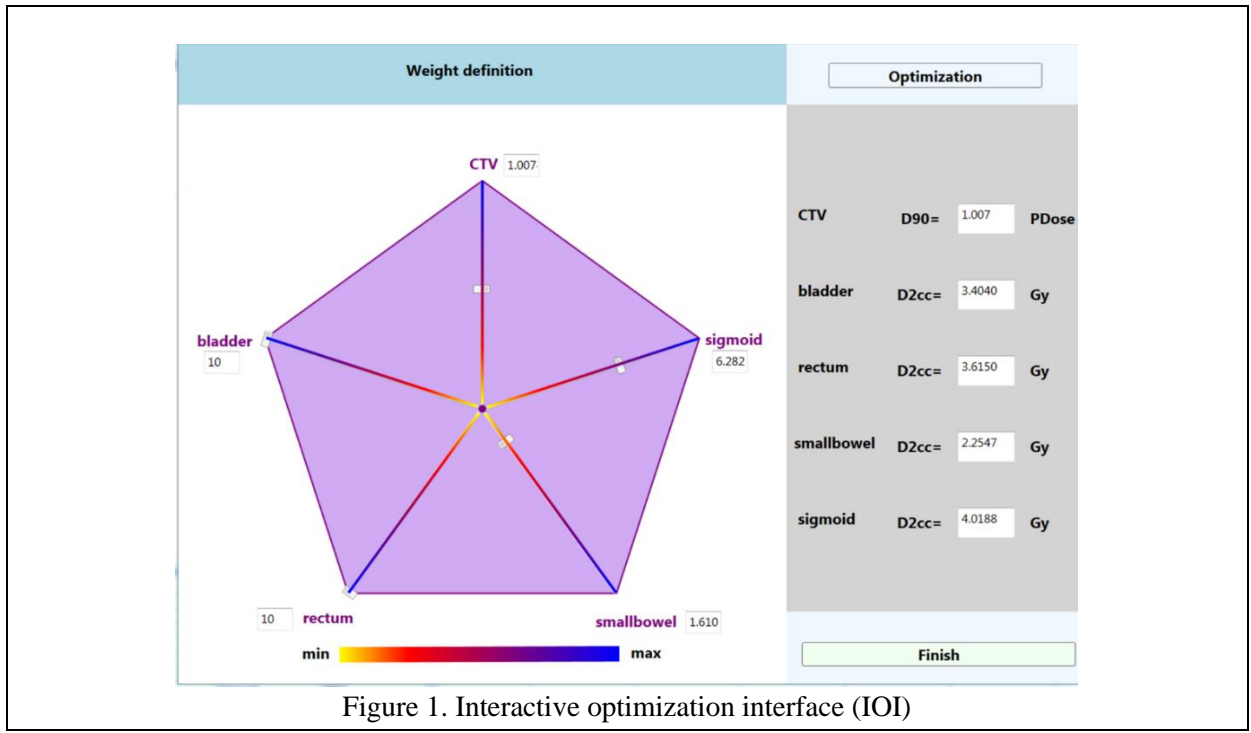
Figure 1. Interactive optimization interface (IOI)

## 2.3 Testing cases

Two clinical examples of seeking dwell time matrix of the source in gynecologic cervix cancer treatment were chosen to test the accuracy and efficiency of IOI. The T&O applicator Fletcher-Suit was used to deliver the source radiation. In the first example (patient A), there are 21 possible dwell positions (resolution of 5 mm in three applicators), and the prescription dose is 580 cGy. In the second example (patient B), there are 22 possible dwell positions (resolution of 5 mm in three applicators), and the prescription dose is 600 cGy. The dose voxel resolution of these two patients is 1.1719 mm ∗ 1.1719 mm ∗ 2 mm. The computational platform is a standard desktop computer with CPU processor Inter(R) Core (TM) i7-9700K CPU @ 3.60GHz and memory 32 GB. The optimization speed can be further improved with parallel computation.

## 3. Results

### 3.1 Tandem and ovoid case

For these two examples, before optimization, the treatment planner needed good protection for the bladder and rectum. In addition, for the purpose of comparison, the value of $D_{90}$ for CTV was set the same as that of the approved clinical plan. During the optimization process, the $D_{90}$ value for patient A and patient B was set to 589.49 cGy (1.016 pdose) and 604.46 cGy (1.090 pdose) respectively, and the weighting factors of OARs were constantly adjusted to make sure the $D_{2cc}$ values for bladder and rectum were as low as possible. In the interactive optimization interface, the treatment planner needed to simply increase the weight for the bladder and rectum and, meanwhile, decrease the weight for the sigmoid and small bowel. When the treatment planner ascertains that the $D_{2cc}$ value for OARs is satisfactory, the optimization is complete.

In order to check the accuracy of IOI, the final matrix of source dwell time obtained was imported into the clinical treatment planning system BrachyVision via DICOM-RT data (re-import plan). The $D_{90}$ value for CTV and $D_{2cc}$ values for OARs and the DVHs for CTV and OARs obtained from IOI were compared with those obtained from the re-import plan in BrachyVision. Table 1 and Table 2 presented $D_{90}$ values and $D_{2cc}$ values for patient A and patient B, respectively. The difference of $D_{90}$ value and $D_{2cc}$ values between the optimization plan and the re-import plan is within 2% for these two patient cases. Figure 2 show the DVH curves for CTV and OARs for patient A and patient B. The DVH curves of the optimization plan (dotted line) almost coincide with those of the re-import plan (solid line). Therefore, the dose results exported from IOI are accurate.

Table 1. Comparison of $D_{90}$ for CTV and $D_{2cc}$ for OARs between the optimization plan and the re-import plan in Eclipse for patient A.

| Items | Target Quantity | Optimization | Eclipse | Different |
|---|---|---|---|---|
| Bladder | $D_{2cc}$ | 4.439 Gy | 4.474 Gy | -0.80% |
| Sigmoid | $D_{2cc}$ | 3.520 Gy | 3.520 Gy | 0.02% |
| Rectum | $D_{2cc}$ | 3.218 Gy | 3.207 Gy | 0.34% |
| Small bowel | $D_{2cc}$ | 2.035 Gy | 2.019 Gy | -0.79% |
| CTV | $D_{90}$ | 589.280 cGy | 592.169 cGy | -0.49% |

Table 2. Comparison of $D_{90}$ for CTV and $D_{2cc}$ for OARs between the optimization plan and the re-import plan in Eclipse for patient B.

| Items | Target Quantity | Optimization | Eclipse | Different |
|---|---|---|---|---|
| Bladder | $D_{2cc}$ | 3.404 Gy | 3.397 Gy | 0.20% |
| Sigmoid | $D_{2cc}$ | 4.019 Gy | 3.947 Gy | 1.80% |
| Rectum | $D_{2cc}$ | 3.615 Gy | 3.660 Gy | -1.20% |
| Small bowel | $D_{2cc}$ | 2.255 Gy | 2.256 Gy | -0.10% |
| CTV | $D_{90}$ | 604.440 cGy | 602.087 cGy | 0.40% |

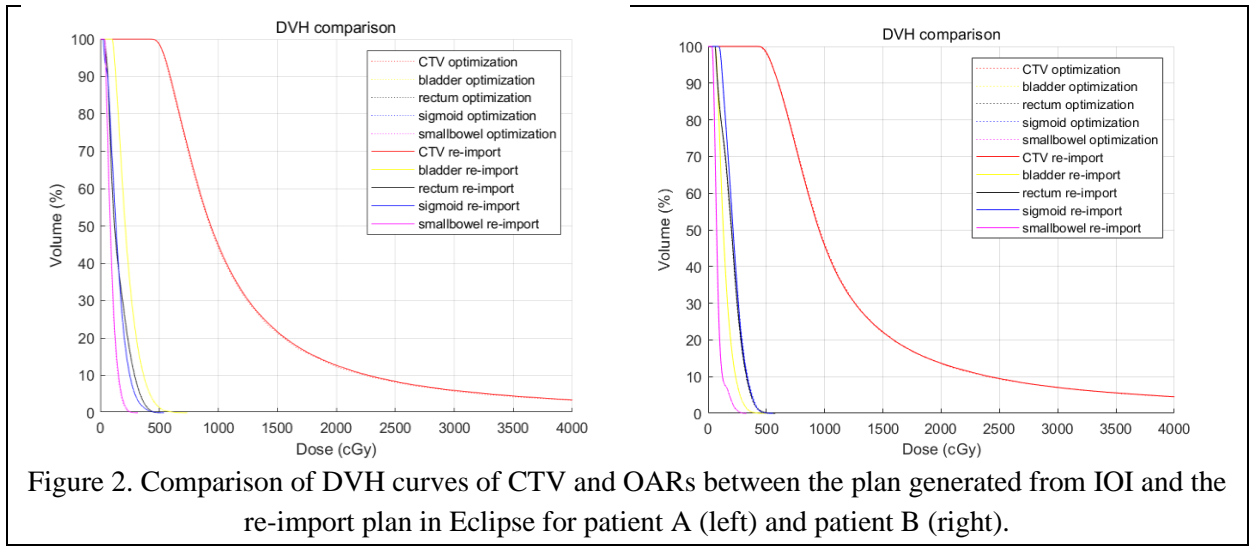

Figure 2. Comparison of DVH curves of CTV and OARs between the plan generated from IOI and the re-import plan in Eclipse for patient A (left) and patient B (right).

*3.2 Comparison with manual planning*

In order to test the efficiency, we compared the optimization plan with the clinically approved plan. In BrachyVision optimization tool, the treatment planner had to vary the weighting factors multiple times to find the optimum solution, and in most cases, did not know how to make modifications. Therefore, it took a long period of time to do the optimization using this procedure. Thus, the BrachyVison optimization tool was not in use in our institute but the treatment planner preferred the trial-and-error manipulation of dwell times and positions for treatment planning.

For comparison, the results of the optimization plan and clinically approved plan for patient A are presented in table 3 and figure 3 (left). When comparing the values in table 3, we can see that the doses to the sigmoid, rectum and small bowel in the optimization plan are much improved, while the dose to the bladder is slightly higher, with identical $D_{90}$ values for CTV. It can be seen in figure 3 (left) that DVH values start to decline obviously faster using the optimization plan. Table 4 and figure 3 (right) demonstrate the results of the optimization plan and the clinically approved plan for patient B. We can see from table 4 that all OAR doses using the optimization plan are smaller than those for the clinically approved plan. Figure 3 (right) clearly shows that DVH values in the optimization plan also decline faster than those in the clinical plan. The time necessary to generate the plan using the IOI and trial-and-error method in BrachyVision is shown in table 5. Overall, compared to the trial-and-error method of manipulating dwell times and positions, the IOI can make treatment planning times shorter and improve the treatment quality of the treatment plan.

Table 3. Comparison of $D_{90}$ for CTV and $D_{2cc}$ for OARs between the clinically approved plan and the optimization plan for patient A

| Items | Target Quantity | Clinical Plan | Optimization Plan | Dose Reduction |
|---|---|---|---|---|
| Bladder | $D_{2cc}$ | 4.372 Gy | 4.439 Gy | 1.532% |
| Sigmoid | $D_{2cc}$ | 3.788 Gy | 3.520 Gy | -7.075% |
| Rectum | $D_{2cc}$ | 3.385 Gy | 3.218 Gy | -4.934% |
| Small bowel | $D_{2cc}$ | 2.150 Gy | 2.035 Gy | -5.349% |
| CTV | $D_{90}$ | 589.486 cGy | 589.280 cGy | -0.035% |

Table 4. Comparison of $D_{90}$ for CTV and $D_{2cc}$ for OARs between the clinically approved plan and the optimization plan for patient B.

| Items | Target Quantity | Clinical Plan | Optimization Plan | Dose Reduction |
|---|---|---|---|---|
| Bladder | $D_{2cc}$ | 3.496 Gy | 3.404 Gy | -2.632% |
| Sigmoid | $D_{2cc}$ | 4.037 Gy | 4.019 Gy | -0.446% |
| Rectum | $D_{2cc}$ | 3.836 Gy | 3.615 Gy | -5.761% |
| Small bowel | $D_{2cc}$ | 2.690 Gy | 2.255 Gy | -16.171% |
| CTV | $D_{90}$ | 604.459 cGy | 604.440 cGy | -0.003% |

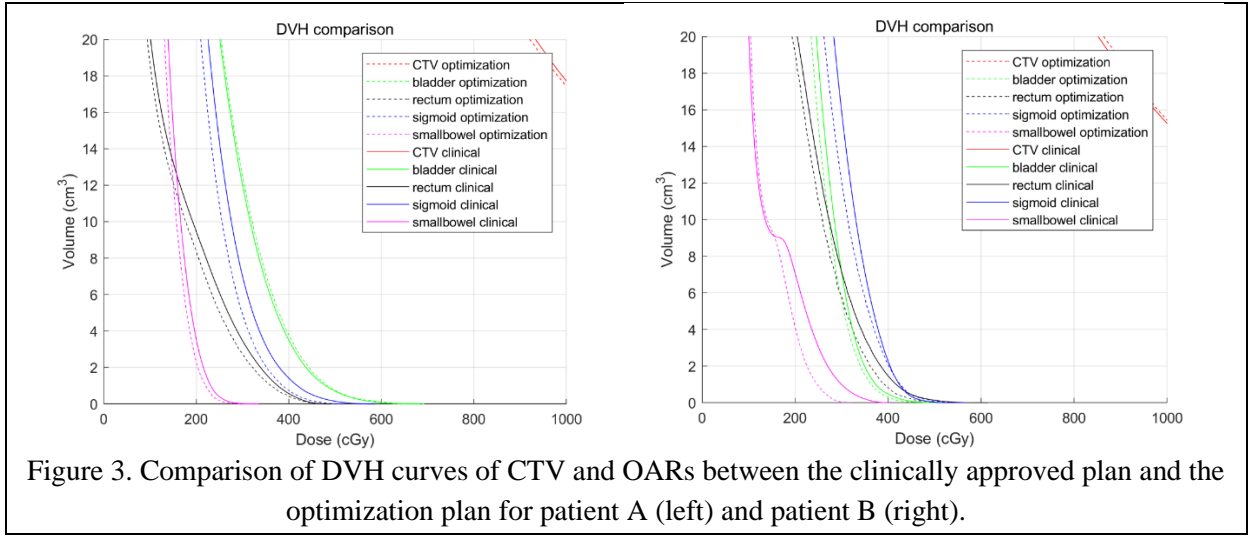
Figure 3. Comparison of DVH curves of CTV and OARs between the clinically approved plan and the optimization plan for patient A (left) and patient B (right).

Table 5. Comparison of CPU time for treatment planning between the clinically approved plan and the optimization plan for both patients.

| Items | Clinical Plan | Optimization Plan |
|---|---|---|
| Patient 1 | 9.5 min | 0.9 min |
| Patient 2 | 9.8 min | 1.0 min |

## 4. Discussions and Conclusions

In this paper, we proposed a novel approach IOI for volumetric HDR brachytherapy optimization. In this approach, both the fact of minimizing the dose for OARs and having a good dose coverage for CTV are considered in the model of optimization, and the treatment planner can search an optimal solution using an interactive method. In the process of optimization, the physicist interactively defines the weighting factors for CTV and OARs according to the physician's preference of protection for OARs. The clinical cared quantities, which include the $D_{90}$ for CTV and $D_{2cc}$ for OARs, will be directly shown in the interactive interface. In this way, the physicist can intuitively judge if this optimization step is satisfactory. We have demonstrated that our interactive optimization method can make treatment planning times shorter and improve the quality of the treatment plans by two examples of treatment planning for gynecological cervix cancer.

It takes about 1 minute to obtain a good quality plan using our method IOI, which is obviously faster than trial-and-error planning. The optimization efficiency can be further improved by adding GPU into our current program.

This interactive optimization program is developed by visual C#, which is compatible with Eclipse Scripting API. We plan to integrate this interactive optimization interface with Eclipse through Eclipse Scripting API. As a result, when a physicist generates the patient geometry and begins the optimization, an API script of our optimization is called to complete the optimization and then the optimal solution for dwell time matrix is automatically imported into eclipse through API function. This integration will smooth the treatment planning process using Eclipse and drastically improve the efficient of treatment planning for HDR brachytherapy.

**Reference**

1   A.N. Viswanathan, J. Moughan, W. Small, C. Levenback, R. Iyer, S. Hymes, A.P. Dicker, B. Miller, B. Erickson, D.K. Gaffney, "The Quality of Cervical Cancer Brachytherapy Implantation and the Impact on Local Recurrence and Disease-Free Survival in Radiation Therapy Oncology Group Prospective Trials 0116 and 0128," Int J Gynecol Cancer **22**, 123-131 (2012).

2   D.G. Petereit, S.J. Frank, A.N. Viswanathan, B. Erickson, P. Eifel, P.L. Nguyen, D.E. Wazer, "Brachytherapy: Where Has It Gone?," J Clin Oncol **33**, 980-+ (2015).

3   J. Mayadev, L. Qi, S. Lentz, S. Benedict, J. Courquin, S. Dieterich, M. Mathai, R. Stern, R. Valicenti, "Implant time and process efficiency for CT-guided high-dose-rate brachytherapy for cervical cancer, Brachytherapy," Brachytherapy **13**, 233-239 (2014).

4   M.S. Bazaraa, H.D. Sherali, C.M. Shetty, *Nonlinear Programming: Theory and Algorithms*. (Hoboken: A John Wiley & Sons, 2006).

5   Y. Gonzalez, C. Shen, P. Klages, K. Albuquerque, X. Jia, "abstract submitted to AAPM annual meeting," 2018).

6   R. Mazeron, P. Maroun, P. Castelnau-Marchand, I. Dumas, E.R. del Campo, K. Cao, A. Slocker-Escarpa, R. M'Bagui, F. Martinetti, A. Tailleur, A. Guemnie-Tafo, P. Morice, C. Chargari, D. Lefkopoulos, C. Haie-Meder, "Pulsed-dose rate image-guided adaptive brachytherapy in cervical cancer: Dose–volume effect relationships for the rectum and bladder," Radiotherapy and Oncology **116**, 226-232 (

7   R. Henri, M. Kaisa, P. Jan-Erik, L. Tapani, "Interactive multiobjective optimization for anatomy-based three-dimensional HDR brachytherapy," Physics in Medicine & Biology **55**, 4703 (2010).

8   K. Miettinen, M.M. Mäkelä, "Interactive bundle-based method for nondifferentiable multiobjeective optimization: nimbus," Optimization **34**, 231-246 (1995).

9   R. Pötter, C. Haie-Meder, E.V. Limbergen, I. Barillot, M.D. Brabandere, J. Dimopoulos, I. Dumas, B. Erickson, S. Lang, A. Nulens, P. Petrow, J. Rownd, C. Kirisits, "Recommendations from gynaecological (GYN) GEC ESTRO working group (II): Concepts and terms in 3D image-based treatment planning in cervix cancer brachytherapy—3D dose volume parameters and aspects of 3D image-based anatomy, radiation physics, radiobiology," Radiotherapy and Oncology **78**, 67-77 (2006).

10  I.C.o.R. Units, Measurements, *Prescribing, Recording, and Reporting Brachytherapy for Cancer of the Cervix*. (Oxford University Press, 2016).

11  Z. Yuhong, K. Peter, T. Jun, C. Yujie, S. Strahinja, Y. Ming, H. Brian, M. Paul, P. Arnold, J. Steve, A. Kevin, J. Xun, "Automated high-dose rate brachytherapy treatment planning for a single-channel vaginal cylinder applicator," Physics in Medicine & Biology **62**, 4361 (2017).

12  F. Shi, B. Hrycushko, J. Tan, P. Medin, S. Stojadinovic, A. Pompos, M. Yang, K. Albuquerque, X. Jia, "An Automated Treatment Plan Quality Assurance Program for Tandem and Ovoid High Dose-Rate Brachytherapy," Brachytherapy **14**, S67 (

13  P. Klages, C. Shen, J. Tan, K. Albuquerque, X. Jia, "Automated HDR Brachytherapy Treatment Planning and Quality Assurance System for Tandem-And-Ovoid Applicator," Medical Physics **44**2017).